\documentclass[preprint,onecolumn]{revtex4-1}
\pdfoutput=1

\usepackage{amssymb,amsfonts,amsmath}
\usepackage{graphicx}

\newcommand\Rey{\mbox{\textit{Re}}}  % Reynolds number
\newcommand\Pra{\mbox{\textit{Pr}}}  % Prandtl number, cf TeX's \Pr product
\newcommand\Ray{\mbox{\textit{Ra}}}  % Rayleigh number
\newcommand\Nus{\mbox{\textit{Nu}}}  % Nusselt number
\newcommand\Lew{\mbox{\textit{Le}}}  % Lewis number
\begin{document}

\title{From convection rolls to finger convection in double-diffusive turbulence}

\author{Yantao Yang}
%\email[Corresponds to:]{yantao.yang@utwente.nl}
\affiliation{Physics of Fluids Group, MESA+ Research Institute,  and J. M. Burgers Centre for Fluid Dynamics,
                    University of Twente, PO Box 217, 7500 AE Enschede, The Netherlands.}

\author{Roberto Verzicco}
\affiliation{Physics of Fluids Group, MESA+ Research Institute,  and J. M. Burgers Centre for Fluid Dynamics,
                    University of Twente, PO Box 217, 7500 AE Enschede, The Netherlands.}
\affiliation{Dipartimento di Ingegneria Industriale, University of Rome ``Tor Vergata'',
                    Via del Politecnico 1, Roma 00133, Italy}
                    
\author{Detlef Lohse}
\affiliation{Physics of Fluids Group, MESA+ Research Institute,  and J. M. Burgers Centre for Fluid Dynamics,
                    University of Twente, PO Box 217, 7500 AE Enschede, The Netherlands.}
\affiliation{Max-Planck Institute for Dynamics and Self-Organization, Am Fassberg 17, 37077 G\"{o}ttingen, Germany.}

\date{\today}

\begin{abstract}
Double diffusive convection (DDC), which is the buoyancy driven flow with fluid density depending on two scalar components, is ubiquitous in many natural and engineering enviroments. Of great interests are scalers transfer rate and flow structures. Here we systematically investigate DDC flow between two horizontal plates, driven by an unstable salinity gradient and stabilized by a temperature gradient. Counterintuitively, when increasing the stabilizing temperature gradient, the salinity flux first increases, even though the velocity monotonically decreases, before it finally breaks down to the purely diffusive value. The enhanced salinity transport is traced back to a transition in the overall flow pattern, namely from large scale convection rolls to well-organised vertically-oriented salt fingers. We also show and explain that the unifying theory of thermal convection originally developed by Grossmann and Lohse for Rayleigh-B\'{e}nard convection can be directly applied to DDC flow for a wide range of control parameters (Lewis number and density ratio), including those which cover the common values relevant for ocean flows.
\end{abstract}

\maketitle

Double diffusive convection (DDC), where the flow density depends on two scalar components, is of great relevance in many natural phenomena and engineering applications, such as oceanography~\cite{Turner1985,Schmitt1994,Schmitt2005}, geophysics~\cite{Schoofs1999,Buffett2010}, astrophysics~\cite{Merryfield1995,Rosenblum2011,Leconte2012,Mirouh2012,Wood2013}, and process technology~\cite{DHernoncourt2006}. A comprehensive review of the field can be found in the recent book of Ref.~\cite{Radko2013}. In DDC flows the two components usually have very different molecular diffusivities. For simplicity and to take the most relevant example, we refer to the fast diffusing scalar as temperature and the other as salinity, but our results are more general. The difference between the diffusing time scales of two components induces interesting flow phenomena, such as the well-known salt fingers observed in ocean flows~\cite{Stern1960,Schmitt2005}. 

In laboratory experiments salt fingers can grow from a sharp interface~\cite{Turner1967} or inside a layer which has uniform scalar gradients and is bounded by two reservoirs~\cite{Linden1978,Krishnamurti2003}. For the latter case a single finger layer or a stack of alternating finger and convection layers was observed for different control parameters. Inside the finger layers long narrow salt fingers develop vertically, while in convection layer fluid is well mixed by large scale circulation. Recent experiments~\cite{Hage_Tilgner2010} revealed that fingers emerge even when the density ratio, i.e., the ratio of the buoyancy force induced by temperature gradient to that by salinity gradient, is smaller than 1. This extends the traditional finger regime where the density ratio is usually larger than 1, and inspired a reexamination of the salt-finger theory which confirmed that salt fingers do grow in this new finger regime~\cite{Schmitt2011}. When the density ratio is small enough, however, finger convection breaks down and gives way to large scale convection rolls, i.e. the flow recovers the Rayleigh-B\'{e}nard (RB) type~\cite{Kellner_Tilgner2014}.

Given the ubiquitousness of DDC in diverse circumstances, it is challenging to experimentally investigate the problem for a wide range of control parameters. Here we conduct a systematic numerical study of DDC flow between two parallel plates which are perpendicular to gravity and separated by a distance $L$. The details of the numerical method is briefly described in the Method section. The top plate has both higher salinity and temperature, meaning that the flow is driven by the salinity difference $\Delta_S$ across the layer and stabilised by temperature difference $\Delta_T$. The molecular diffusivity $\lambda_\zeta$ of a scalar component is usually measured by its ratio to the kinematic viscosity $\nu$, i.e.~the Prandtl number $\Pra_\zeta=\nu / \lambda_\zeta$. Hereafter $\zeta=T$ or $S$ denotes the quantity related to temperature or salinity. The strength of the driving force is measured by the Rayleigh number $\Ray_\zeta=(g \beta_\zeta \Delta_\zeta L^3) / (\lambda_\zeta \nu)$ with $g$ being the gravitational acceleration and $\beta_\zeta$ the positive expansion coefficient. The relative strength of the buoyancy force induced by temperature difference compared to that induced by salinity difference is measured by density ratio defined as $\Lambda=(\beta_T\Delta_T)/(\beta_S\Delta_S)=\Lew\,\Ray_T\,\Ray^{-1}_S$. When $\Lambda=0$ the flow is of RB type and purely driven by the salinity difference. $\Lambda<1$ ($>1$) corresponds to an overall unstable (stable) stratification. Linear stability analysis revealed that instabilities occur as long as $\Lambda<\Lew$~\cite{Stern1960}. As we will show below, the direct numerical simulations of the fully non-linear system indicate that flows develop in the same parameter range, i.e. $\Lambda<\Lew$.

Previous experiments with a heat-copper-ion system~\cite{Kellner_Tilgner2014} showed that as $\Lew$ increases from zero, the flow transits from large convective rolls to salt fingers, which is accompanied by an increase of the salinity transfer. However, the experiments were conducted with a single type of fluid and thus only one combination of Prandtl numbers was investigated. Moreover, the highest density ratio realised in experiments was of order 1. In the present study we will take advantage of numerical simulations which can be easily carried out for a wide range of Prandtl numbers and allow for a more systematic investigation of the problem. We set $\Pra_T=7$, which is the typical value for seawater at $20\,^{\circ}\mathrm{C}$. Several sets of simulations are conducted with different $\Pra_S$ and $\Ray_S$. Since $\Pra_T$ is fixed for all simulations, we can alternatively use the Lewis number $\Lew=\lambda_T/\lambda_S=\Pra_S/\Pra_T$ and $\Ray_S$ to label different sets. Specifically, we run five sets with $(\Lew,\Ray_S)=(1,10^8)$, $(10,10^8)$, $(100,10^7)$, $(100,10^8)$, and $(100,10^9)$, respectively. Within each set we gradually increase $\Lambda$ from 0 (i.e. RB flow) to a value very close to $\Lew$.

\begin{figure}
\centerline{\includegraphics[scale=1.0]{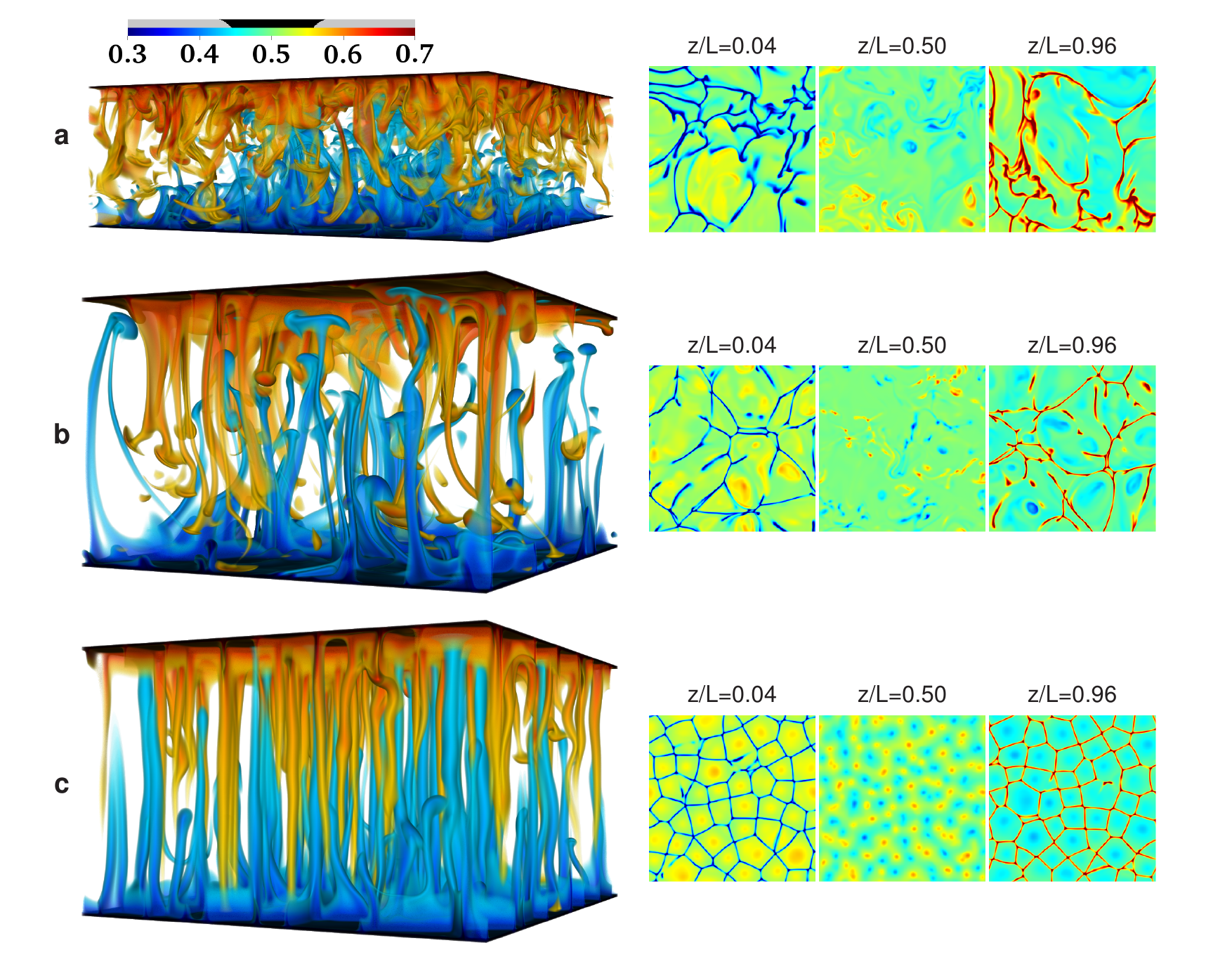}}%
\caption{Different types of flow structures observed in simulations with (a) $(\Lew, \Lambda)=(1, 0.1)$, (b) $(\Lew, \Lambda)=(100, 10^{-3})$, and (c) $(\Lew, \Lambda)=(100, 1)$. For all three cases $\Ray_S=10^8$. We show the three dimensional (3D) rendering of structures with low (blue) and high (red) salinity, and salinity contours on three horizontal slices at different heights. The same colormap is used for all plots. In the 3D plots the opacity is also set by salinity, as indicated by the legend. In (a) the plumes gather into clusters and move collectively in vertical direction, which drives the large scale convection rolls. In (b) the plumes becomes thinner due to the larger $\Pra_S$, but they still form clusters and large scale convection rolls. In (c) the large-scale rolls are replaced by well-organised vertically-oriented salinity fingers, which extend through the entire domain heights. In all 3D plots the saltier and fresher plumes (or fingers) develop from top and bottom plates, respectively.}
\label{fig:finger}
\end{figure}

In Fig.~\ref{fig:finger} we show the typical flow structures observed in our simulations. For $\Lew=1$, even with $\Lambda$ up to $0.1$ as shown in Fig.~\ref{fig:finger}a the flow structures are very similar to those in the RB case. Near boundaries sheet structures emerge as the roots of salt plumes, e.g.~see the contours on two slices at $z=0.04$ and $0.96$ in Fig.~\ref{fig:finger}a. These sheet structures gather in some regions, from where the salt plumes emit into the bulk as clusters. The plume clusters move collectively and drive the large-scale convection rolls. When $\Lew>1$, flow structures are of RB type at small $\Lambda$, as shown in Fig.~\ref{fig:finger}b. The flow morphology is essentially the same as in Fig.~\ref{fig:finger}a, i.e. the salt plumes still form clusters and drive the large scale rolls. The salt plumes become thinner and more circular due to the larger $\Pra_S$ than that in Fig.~\ref{fig:finger}a. At moderate $\Lambda=1.0$, however, the salt plumes stop gathering and convection rolls are replaced by vertically-oriented salt fingers. The highly organising pattern can be found both in the sheet structures near plates and the salt fingers in the middle, as indicated by contours on three slices shown in Fig.~\ref{fig:finger}c. These well-organised fingers develop separately and extend from one plate to the opposite one. When $\Lambda$ increases close to $\Lew$, all flow motions are suppressed by the strong temperature field for all $\Lew$'s considered here.

\begin{figure}
\centerline{\includegraphics[scale=1.0]{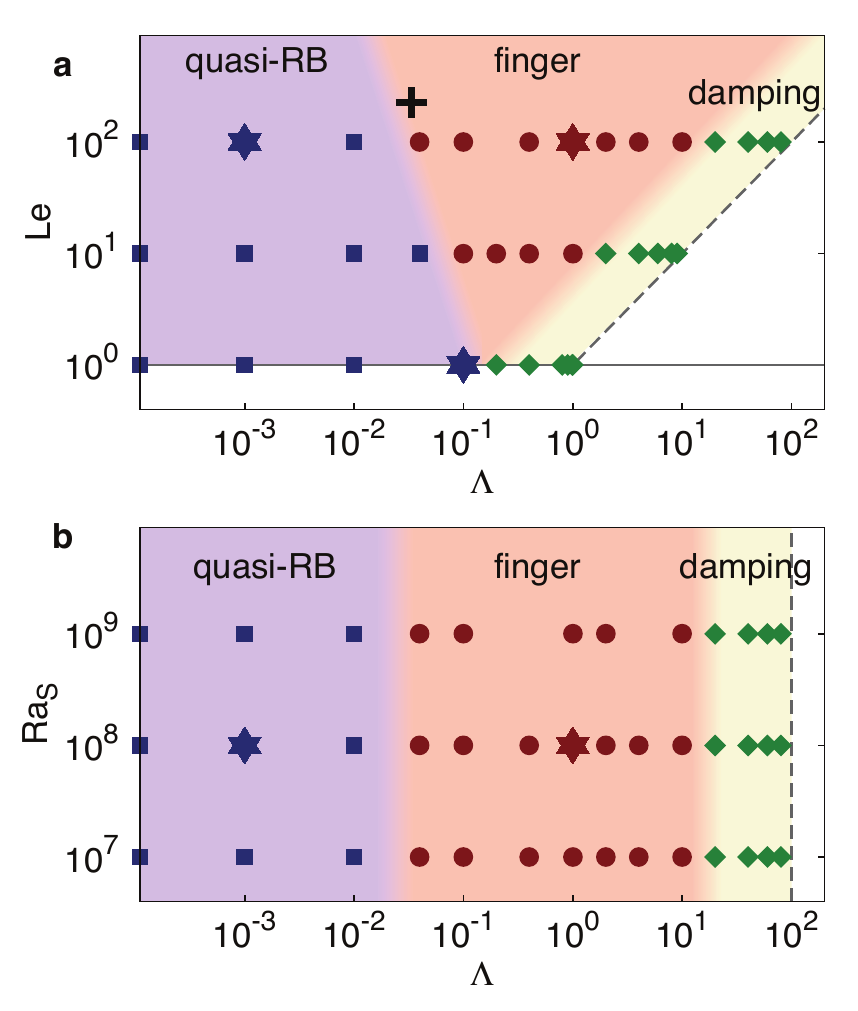}}%
\caption{Explored phase space and schematic illustration of different flow regimes. (a) The three sets of simulations with $\Ray_S=10^8$ are shown in the $\Lambda-\Lew$ plane, and (b) the three sets with $\Lew=100$ are shown in the $\Lambda-\Ray_S$ plane, respectively. The top row in (a) and the middle row in (b) correspond to the same set of simulations. The horizontal solid line in (a) marks $\Lew=1$, below which the flow enters the diffusive regime of DDC, i.e. the fast diffusing component drives the flow. The dashed lines in both panels represent the stability limit $\Lambda=\Lew$. Three flow regimes can be identified and indicated by different colours: The quasi-RB regime (blue), the finger regime (orange), and the damping regime (grey). The three stars in (a) and two stars in (b) mark the cases shown in Fig.~\ref{fig:finger}. The black plus sign in (a) indicate the transition point reported in Ref.~\cite{Kellner_Tilgner2014}.}
\label{fig:pd}
\end{figure}

Based on the flow morphology observed in simulations, different flow regimes can be identified. In Fig.~\ref{fig:pd} we present the explored control parameters and a schematic division of phase space into three regimes based on the numerical observations. The three sets with the same $\Ray_S$ and different $\Lew$ are shown in the $\Lambda$-$\Lew$ phase plane, see Fig.~\ref{fig:pd}a. For very small density ratio the flow is dominated by large-scale convection rolls, which we refer to as the quasi-RB regime. When $\Lambda$ is very close to $\Lew$ all flow motions start to be suppressed by the strong temperature field, which we refer to as the damping regime. When $\Lew=1$ the flow directly transits from the quasi-RB regime into the damping regime as $\Lambda$ increases. For $\Lew>1$ salt fingers develop at moderate $\Lambda$ and a finger regime can be identified. As $\Lew$ increases the finger regime occupies a wider range of $\Lambda$. The transition point between the quasi-RB and the finger regime for the heat-copper-ion system has been experimentally determined at $(\Lambda,\,\Lew)\approx(226, \, 1/30)$~\cite{Kellner_Tilgner2014}, which is also marked in Fig.~\ref{fig:pd}a, and it is very close to the transition boundary found in the current study. For fixed $\Lew=100$, the transition between regimes happens at similar $\Lambda$ for different $\Ray_S$. Similar behaviour of the transition between the quasi-RB and finger regimes has been discovered experimentally for $\Lew\approx226$~\cite{Kellner_Tilgner2014}, i.e. the transition is independent of $\Ray_S$. However, in the experiment the highest density ratio is of order 1 and therefore only the quasi-RB and finger regimes were identified~\cite{Kellner_Tilgner2014}.

In Ref.~\cite{Kellner_Tilgner2014} the authors proposed two possible scaling laws to describe the transition between the quasi-RB and finger regimes, i.e. $\Lambda=const.$ or $\Ray_T\sim\Pra_T^{6/7}\Ray_S^{22/21}$. The latter one is equavlent to $\Lambda\sim\Pra_T^{6/7}\Ray_S^{1/21}\Lew$. Since all of their experiments have similar $\Pra_T$ and $\Lew$, the only difference between the two possibilities is the factor $\Ray_S^{1/21}$ with an exponent too small to be distinguished by the experimental measurement. However, the two scalings have different dependences on $\Lew$, which can be tested against our numerical results. From Fig.~\ref{fig:pd}a one observes that as $\Lew$ increases the transition to the finger regime happens at smaller $\Lambda$, which contradicts the second scaling. However, the current results are compatible with the first scaling.

\begin{figure}
\centerline{\includegraphics[scale=1.0]{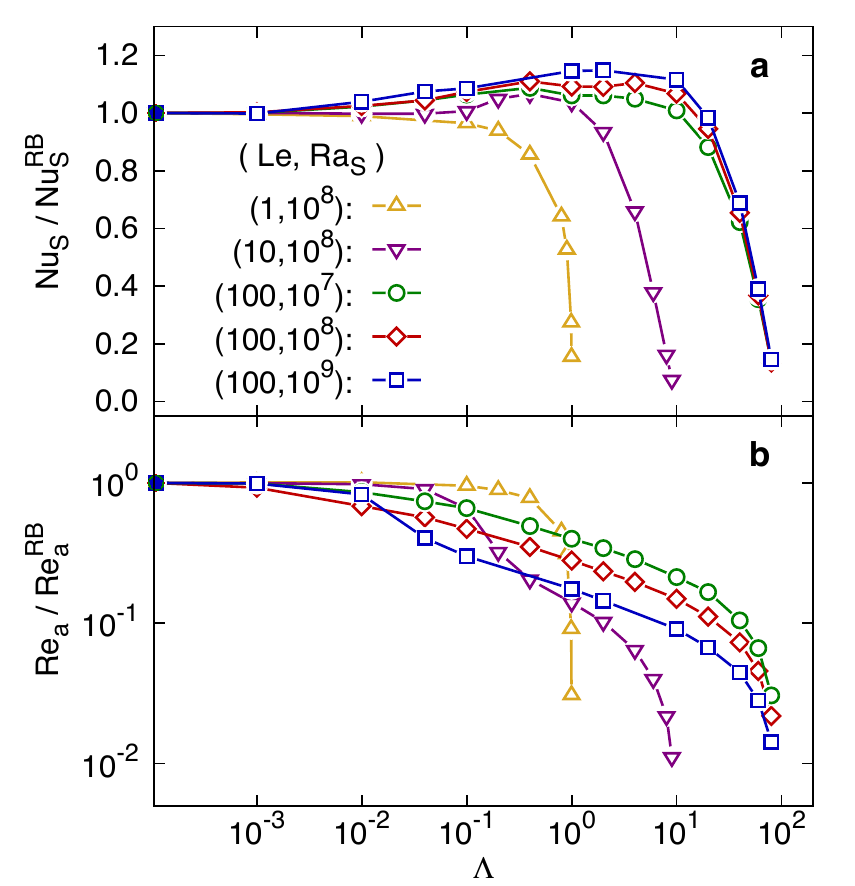}}%
\caption{(a) Salinity flux $\Nus_S$ and (b) Reynolds number $\Rey_a$ versus density ratio $\Lambda$ for different Lewis numbers and Rayleigh numbers. All quantities are normalized by the values of Rayleigh-B\'{e}nard (RB) flow with the same $\Ray_S$ and $\Pra_S$. The solid symbols on the vertical axes represent the RB cases within each set. $\Rey_a$ decreases monotonically for all sets. But $\Nus_S$ can be larger than $\Nus^{RB}_S$ in the finger regime at $\Lew>1$.}
\label{fig:response}
\end{figure}

Different flow structures have significant influences on the global responses of system. The two most important responses are the salinity flux and the flow velocity, which are usually measured by the Nusselt number $\Nus_S$ and the Reynolds number $\Rey_a$.
\begin{equation}
  \Nus_S=\frac{\langle u_3 s \rangle - \lambda_S\partial_3\langle s \rangle}{\lambda_S \Delta_S L^{-1}},
  \quad \Rey_a=\frac{u_{rms} L}{\nu}.
\end{equation}
Here $u_3$ is vertical velocity, $s$ is salinity, $\partial_3$ is vertical derivative, $\langle\cdot\rangle$ is the average over time and the entire domain, and $u_{rms}$ is the rms value of velocity magnitude, respectively. In Fig.~\ref{fig:response} we plot the variations of $\Nus_S$ and $\Rey_a$ normalized by the values of corresponding RB flow (denoted by superscript ``RB'') as $\Lambda$ increases from zero to $\Lew$. The two quantities exhibit totally different behaviours in the three regimes. In the quasi-RB regime at small $\Lambda$ both $\Nus_S$ and $\Rey_a$ are very close to $\Nus_S^{RB}$ and $\Rey_a^{RB}$. As $\Lambda$ increases, for the four sets with $\Lew>1$ $\Nus_S$ is larger than $\Nus^{RB}_S$ although $\Rey_a$ decreases according to some effective power-law scaling, which corresponds to the finger regime. When $\Lambda$ becomes large enough and close to $\Lew$, the flow enters the damping regime and both $\Nus_S$ and $\Rey_a$ quickly drop to the values of purely conductive case. For the set with $\Lew=1$ the flow directly transits from the quasi-RB regime to the damping regime, thus no increment of $\Nus_S$ is found in the whole range of $0<\Lambda<1$.

The enhancement of $\Nus_S$ in the finger regime is remarkable since we apply a stabilizing temperature field, but nonetheless the salinity transfer is enhanced. Furthermore, the regime with higher salinity flux extends to $\Lambda>1$ for large $\Lew$. Recall that $\Lambda>1$ corresponds to an overall stable stratification of the fluid. Our results suggest that salinity flux in a stably stratified fluid can exceed that in a unstably stratified state such as purely RB case! For fixed $\Lew$, the increment of $\Nus_S$ is more pronounced at higher $\Ray_S$. The highest increment achieved is about $15\%$, which is comparable to what was found in experiments~\cite{Kellner_Tilgner2014}. However, in our simulations $\Nus_S$ follows a trend which is different from the experiment. In experiments $\Nus_S$ reaches a maximum at the transition from the quasi-RB to the finger regime, while our results indicate that $\Nus_S$ is largest at not the transition but a bigger $\Lambda$. The three sets of simulations at $\Lew=100$ even suggest that there may exit a range of $\Lambda$ in the finger regime where $\Nus_S$ is nearly constant and larger than the RB value. To clarify this discrepancy more simulations are needed at control parameters similar to those in experiments.

\begin{figure}
\centerline{\includegraphics[scale=1.0]{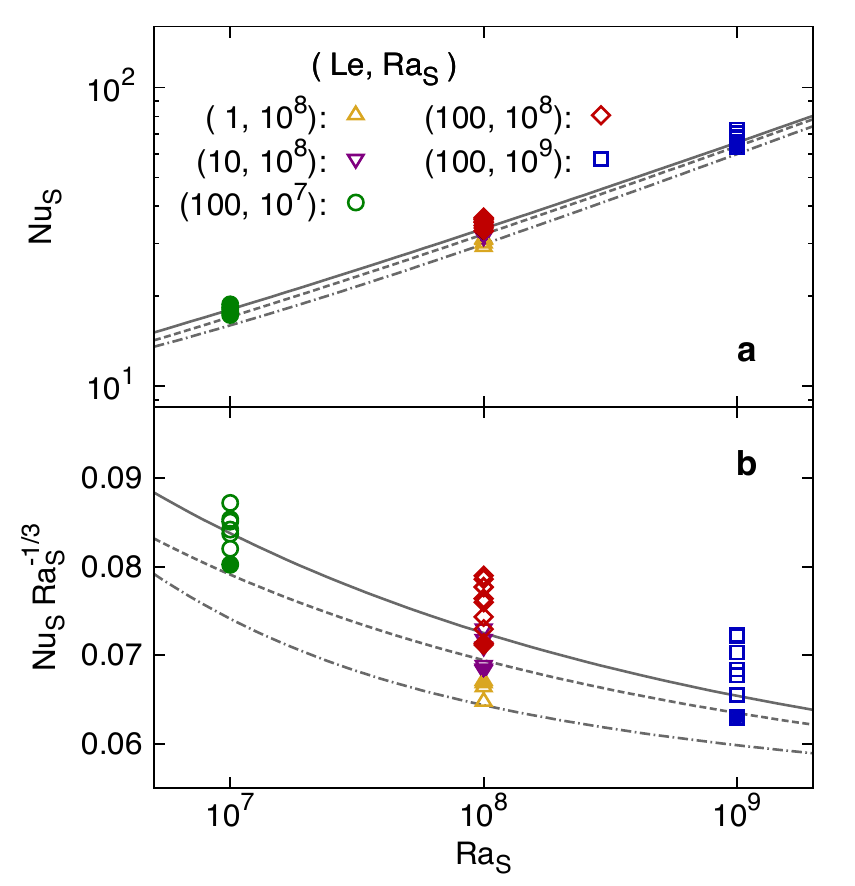}}%
\caption{Comparison between numerical results and the Grossmann-Lohse (GL) theory in their original values (a) and in a compensated way (b). Good agreement can be found between the salinity flux and the GL theory in the quasi-RB and the finger regimes. The GL predictions are shown by the solid line for $\Pra_S=7$, the dashed line for $\Pra_S=70$, and the dash-dotted line for $\Pra_S=700$, respectively.}
\label{fig:gl}
\end{figure}

Our previous study~\cite{yang2015} suggested that the Grossmann-Lohse (GL) model originally developed for RB flow~\cite{GL2000, GL2001, GL2002, GL2004, Stevens_etal2013} can be directly applied to vertically bounded DDC flow. The prediction of the GL model is consistent with both the numerical data~\cite{yang2015} with $\Lew=100$ and $\Lambda\in(0.1,10)$, and the experimental data~\cite{Hage_Tilgner2010} with $\Lew\approx200$ and $\Lambda$ smaller than or close to $1$. Current results indicate that in the quasi-RB regime $\Nus_S$ is almost the same as $\Nus^{RB}_S$, and in the finger regime $\Nus_S$ is slightly higher than but still quite close to $\Nus^{RB}_S$, thus the GL theory should give good prediction of $\Nus_S$ in those two regimes. The largest increment is about $15\%$ for $\Lew=100$ and $\Ray_S=10^9$. The Reynolds number, on the other hand, decreases monotonically towards zero as $\Lambda$ varies from $0$ to $\Lew$, thus it cannot be predicted by the original GL model. The current numerical results are compared to the GL model for salinity transfer by using the same coefficients as in the pure RB problem~\cite{Stevens_etal2013,yang2015}, see Fig.~\ref{fig:gl}. Only the data in the quasi-RB and finger regimes are included. Note that the GL model is used to predict $\Nus_S$ for three different $\Pra_S$ values. Indeed the GL model is quite accurate even when shown in the compensated form, which supports our statement that the GL model can be applied to DDC flow, provided that the flow is in the quasi-RB or the finger regime.

\begin{figure}
\centerline{\includegraphics[scale=1.0]{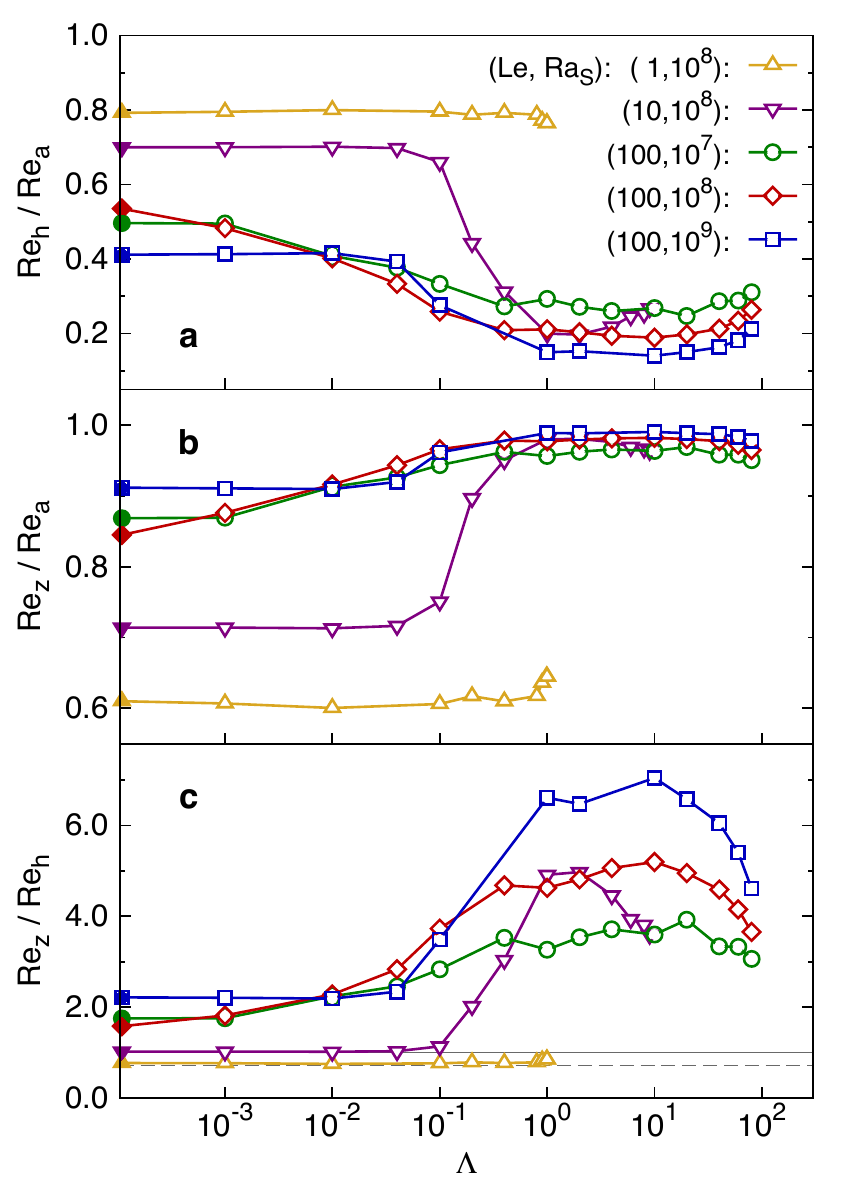}}%
\caption{The ratios between different Reynolds numbers. (a) The ratio $\Rey_h/\Rey_a$ between the Reynolds numbers based on the horizontal velocity and the total velocity, (b) the ratio $\Rey_z/\Rey_a$ between the Reynolds numbers based on vertical velocity and total velocity, and (c) the ratio between $\Rey_z/\Rey_h$. In (c) the horizontal dashed line marks the isotropic value of $\Rey_z/\Rey_h=1/\sqrt{2}$ and the horizontal solid line $\Rey_z/\Rey_h=1$, respectively. The onset of the finger regime is clearly visible by the breakdown of the horizontal velocity and the increase of the vertical one, or the sudden increase of $\Rey_z/\Rey_h$ as shown in (c).}
\label{fig:URt}
\end{figure}

The change of flow morphology can be understood by examining the horizontal and vertical velocities separately. Therefore we define a Reynolds number $\Rey_h$ based on the rms value of the horizontal velocity and a Reynolds number $\Rey_z$ based on the rms value of the vertical velocity. Similar to Ref.~\cite{Kellner_Tilgner2014} we calculate the ratios of $\Rey_h$ and $\Rey_z$ to $\Rey_a$, i.e. the ratios of the horizontal and vertical velocities to the total velocity, see Fig.~\ref{fig:URt}. For $\Lew=1$ both ratios are nearly constant even for $\Lambda$ very close to $\Lew$. Since $\Rey_a$ decreases monotonically to zero as $\Lambda$ approaches $\Lew$, the two curves imply that the stabilizing temperature field damps the horizontal and vertical motions simultaneously. When $\Lew>1$, however, the two ratios follow opposite trends. $\Rey_h/\Rey_a$ and $\Rey_z/\Rey_a$ are constant in the quasi-RB regime with small $\Lambda$. When $\Lambda$ further increases the former decreases to as low as $0.1$ and the latter increases to almost $1$, implying that the fluid moves mainly in the vertical direction and therefore transfers salinity more efficiently. The domination of vertical velocity marks the unset of the finger regime.

We also show in Fig.~\ref{fig:URt}c the ratio $\Rey_z/\Rey_h$, i.e. the ratio of vertical velocity rms to the horizontal velocity rms. For a isotropic flow this ratio should be $1/\sqrt{2}$. When the vertical and horizontal motions are in balance the ratio is $1$. Fig.~\ref{fig:URt}c indicates that in the quasi-RB regime the ratio increases from the isotropic value as $\Lew$ becomes larger. That is, in our numerical simulations the vertical motion is already stronger than the horizontal one for quasi-RB flows at large $\Pra_S$. This is different from the experimental results~\cite{Kellner_Tilgner2014}, where for a much higher $\Pra_S$ the flow is still isotropic in the quatis-RB regime. One possible reason may be the different boundary conditions at the side walls. In our simulations periodic boundary conditions are used for two horizontal directions, while in experiments the side walls are solid and no-slip. Those different horizontal boundary conditions may impose different constrains to the horizontal motions. Nevertheless, for all four sets with $\Lew>1$, the ratio $\Rey_z/\Rey_h$ experiences a sudden increase at the transition from the quasi-RB to the finger regime. This observation is consistent with experimental results~\cite{Kellner_Tilgner2014}, i.e. the transition can be described as a bifurcation.

The results reported here not only reveal some fascinating features about DDC flow for a wide range of control parameters, but also have great application potentials. For instance, for seawater with $\Lew\approx100$ we show that GL model is applicable for $\Lambda$ up to 10, which covers the common value observed in the main thermocline of the subtropical gyres~\cite{Schmitt1994}. Next, transferring scalar component more efficiently in a solution is often desirable in many practical applications. Our results suggest that this can be achieved for a wide range of control parameters, though counterintuitively, by applying a stabilizing thermal gradient to the system. Such enhancement of scalar transfer has been observed in an electrodeposition cell~\cite{Kellner_Tilgner2014}.

\bigskip

\appendix

\section{Methods}
We consider an incompressible flow where the fluid density depends on two scalar component and employ the Oberbeck-Boussinesq approximation, i.e.~$\rho(\theta, s) = \rho_0 [1 - \beta_T \theta + \beta_S s]$. Here $\rho$ is the fluid density, $\rho_0$ is a reference density, $\theta$ and $s$ are the temperature and salinity relative to some reference values, and $\beta_\zeta$ with $\zeta=T$ or $S$ is the positive expansion coefficient associated to scalar $\zeta$, respectively. The flow quantities include three velocity components $u_i$ with $i=1,2,3$, the pressure $p$, and two scalars $\theta$ and $s$. The governing equations read
\begin{subequations}\label{eq:ddc}
\begin{eqnarray}
  \partial_t u_i + u_j \partial_j u_i &=& 
       - \partial_i p + \nu \partial_j^2 u_i + g_i (\beta_T \theta - \beta_S s), \label{eq:mom}  \\
  \partial_t \theta + u_j \partial_j \theta &=& \kappa_T \partial_j^2 \theta,  \\
  \partial_t s + u_j \partial_j s &=& \kappa_S \partial_j^2 s, \label{eq:sal}
\end{eqnarray}
\end{subequations}
where $\nu$ is the kinematic viscosity, $g_i$ is the constant acceleration of gravity, and $\kappa_\zeta$ is the diffusivity of scalar $\zeta$, respectively. The dynamic system is further constrained by the continuity equation $\partial_i u_i = 0$. Without loss of generality, we set $g_1=g_2=0$ and $g_3=g$.

The flow is vertically bounded by two parallel plates separated by a distance $L$. The plates are perpendicular to the direction of gravity. At two plates the no-slip boundary condition is applied, i.e. $u_i\equiv0$, and both scalars are kept constant. The top plate has higher temperature and salinity, thus the flow is driven by the salinity difference $\Delta_S$ across two plates and stabilized by the temperature difference $\Delta_T$. In the two horizontal directions we employ the periodic boundary condition. The horizontal box size is set to be much larger than the horizontal length scales of the flow structures. Initially velocity is set at zero, temperature has a vertically linear profile, and salinity is uniform and equal to the average of boundary values at two plates. The initial fields are similar to those in experiments~\cite{Hage_Tilgner2010}. In order to accelerate the flow development, random noise with a relative amplitude of $0.1\%$ is added to temperature and salinity field. Such initial conditions are used in all simulations.

Equation~\ref{eq:ddc} is nondimensionalized by using the length $L$, the free-fall velocity $U=\sqrt{g\beta_S\Delta_SL}$,  and the scalar differences $\Delta_T$ and $\Delta_S$. To numerically solve the equations we utilised a finite difference solver~\cite{verzicco1996} together with a highly efficient multi-resolution technique~\cite{multigrid2015}. The numerical method has been validated by one-to-one comparisons with experimental results~\cite{yang2015}. 

\begin{acknowledgments}
This study is supported by Stichting FOM and the National Computing Facilities, both sponsored by NWO, the Netherlands. The simulations were conducted on the Dutch supercomputer Cartesius at SURFsara.
\end{acknowledgments}

\end{document}